\documentclass[a4paper,11pt]{article}
\pdfoutput=1 

\usepackage{jinstpub} 
\usepackage{indentfirst}
\title{Proposal on Application of the Multi-Wire Proportional Chambers of the LHCb MUON Detector at Very High Rates for the Future Upgrades}

\author[a,b]{N. Bondar,}
\author[a]{D. Ilin,}
\author[a,1]{O. Maev\note{Corresponding author.}}

\affiliation[a]{Petersburg Nuclear Physics Institute NRC Kurchatov Institute (PNPI NRC KI), Gatchina, Russia}
\affiliation[b]{European Organization for Nuclear Research (CERN), Geneva, Switzerland}

\emailAdd{Oleg.Maev@cern.ch}

\abstract{
The MUON Detector (MD) of LHCb is one of the largest instruments of this kind worldwide, and one of the most irradiated. It has performed exceptionally well during the RUN1 and RUN2 of the LHC at an instantaneous luminosity of 4$\times$10$^{32}$ cm$^{-2}$s$^{-1}$, with tracking inefficiencies at the level of 1$\%$ and 2.6$\%$, respectively.

\setlength{\parskip}{0em}

\par Looking forward for the future LHCb Upgrade 2 (U2) planned in 2031 and aiming in running the detector at increased luminosity by factor $\sim$50, and at the same time keeping a very high ($\sim$99$\%$) detection efficiency, an option with reuse significant part of the present Multi Wire Proportional Chambers (MWPC) in a new Muon System is presented. In addition, the first idea of new Front End Electronics (FEE) and an existing test setup applicable for designing both: new MWPCs with higher granularity of the cathode readout pads and new FEE are described.
}

\keywords{Muon spectrometers; Wire chambers (MWPC); Radiation damage to detector materials (gas detectors).}

\collaboration[c]{on behalf of the LHCb MUON group}

\begin{document}
\maketitle
\flushbottom

\section{Introduction}

\par
\begin{figure}[htbp]
\centering 
\includegraphics[width=.8\textwidth]{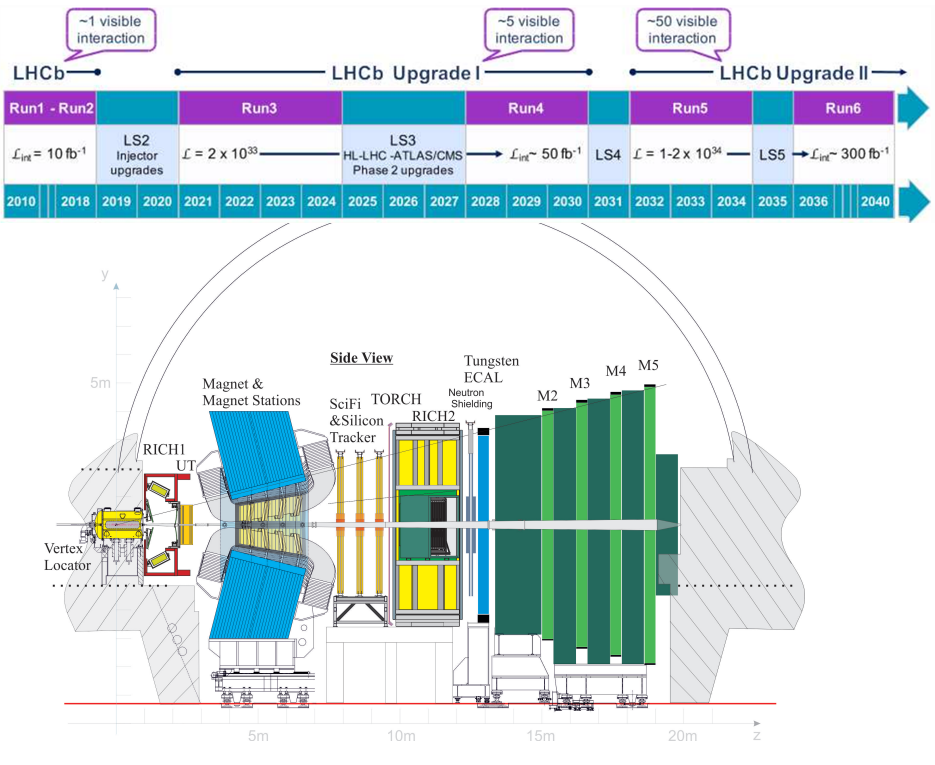}
\caption{\label{fig:1} LHCb long-term operation schedule and schematic side-view of the Upgrade II detector. Muon stations M2--M5 are placed behind the shielding iron wall after electromagnetic calorimeter ECAL and are interleaved with iron absorbers to select penetrating muons.}
\end{figure}

 LHCb experiment \cite{1} is dedicated to look for indirect evidence of new physics in CP violation and rare decays of beauty and charm hadrons. This requires the LHCb detectors (see Fig.\ref{fig:1}), to operate in a very challenging radiation environment, as a results of their forward coverage, with a pseudorapidity range from 2 to 5. Many of the important physics channels are identified by their very clean muon signatures, so the performance of the Muon System is extremely important. To obtain this ambitious goal the MD has been design and successfully operated during RUN1 and RUN2 with a detection efficiency close to 100$\%$ and minor tracking inefficiency. This requirement should be safely forwarded for operation at Upgrade I (U1) and Upgrade II (U2) conditions \cite{2}, see LHCb schedule on Fig.\ref{fig:1}, where instantaneous luminosity will be increased by factor 5 and by a factor 50 from present one $L = 4$$\times$$10^{32}$ cm$^{-2}$s$^{-1}$, respectively. 

\par At present, LHCb considers several scenarios of the new MD for U2 with different type of detectors and experimental techniques \cite{3}. In this document, a proposal with MWPCs as detectors for covering largest area of MD is presented. The very inner part of MD, where the rates are expected such a high as not acceptable for MWPC, it should be instrumented with detectors of another type, in particular using the $\mu$-RWELL technology \cite{4}.  

\section{The LHCb MUON detector}

\par
\begin{figure}[htbp]
\centering 
\includegraphics[width=.9\textwidth]{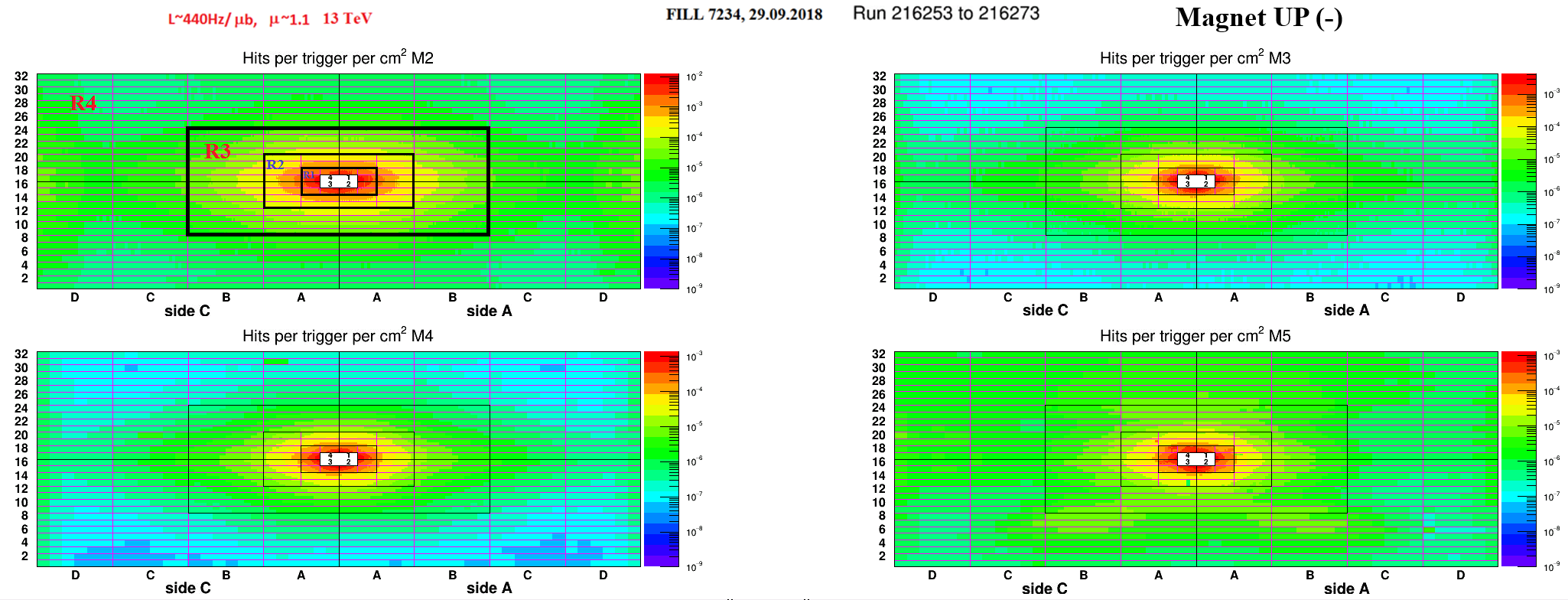}
\caption{\label{fig:2} Occupancy of the hits per trigger per cm$^2$ in M2--M5 stations of the MD from data taken at the end of RUN2. Stations are separated in two halves: C (cryogenics) -- left and A (access) -- right sides, the beam pipe window is in the middle. Each side is segmented in four columns: A,B,C and D and 32 rows with equivalent granularity. On the top left plot the definition of the four regions for station M2 is shown. It is the same for the other stations.}
\end{figure}

 Before U1, the five stations of the MD \cite{1,5} comprised of 1368 MWPCs for a total area of 435 m$^2$. Now, the first station M1 has been removed because of very high expected occupancy, so the MD consists of four stations (M2--M5), for a total area of 387 m$^2$ covered with 1104 MWPCs of 16 different types (see Fig.\ref{fig:2}). Each station is divided into four regions R1--R4, where each region also associated with particular type of MWPC. The area of these four regions scales, from R1 to R4, with the ratios 1 : 4 : 16 : 64, while the irradiation per unit area decreases. This logic has been implemented to provide a hit map for the hardware Level 0 muon trigger during operation in RUN1 and RUN2. Now, at U1 it will be replaced by fully software muon trigger \cite{6}, which makes possible a flexible granularity for the regions and for the physical readout of the MWPCs, following a real occupancy in the MD (which is not homogeneous within regions, see Fig.\ref{fig:2}). Actually, the transition to software trigger is one of the key conditions for this proposal of instrumenting the MD with MWPCs at U2.

\par
\begin{figure}[htbp]
\centering 
\includegraphics[width=1.0\textwidth]{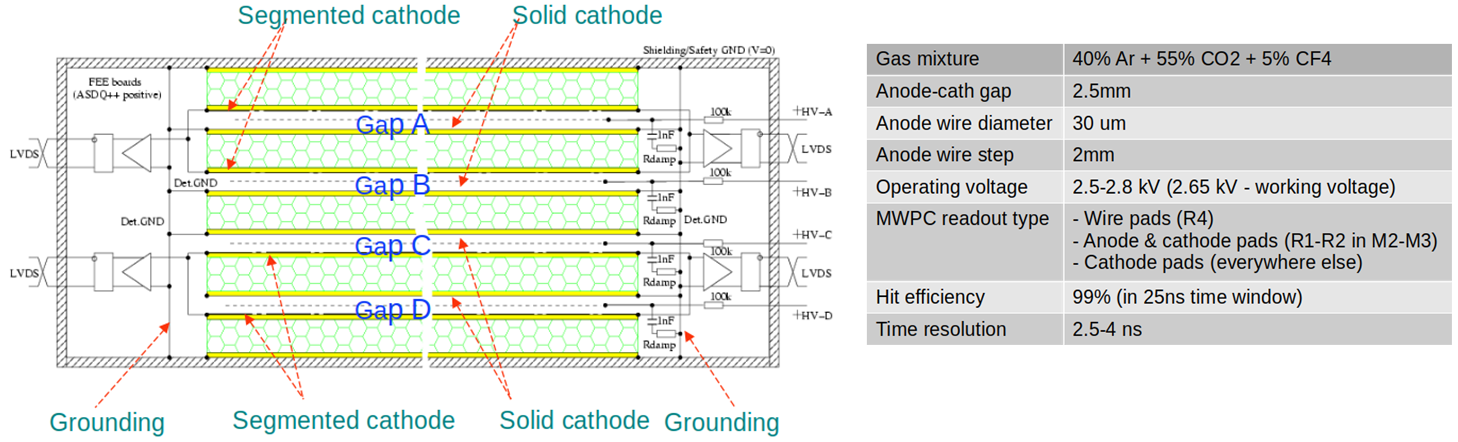}
\caption{\label{fig:3} Cross section of a MWPC with the four gaps indicated by A, B, C and D jointed in the common gas circuit, is shown on the left. Each gap is supplied with an independent HV-line. Corresponding readout pads in contiguous layers (bi-gaps) A$\&$B and C$\&$D are galvanically OR-ed. The most significant parameters of MWPC are presented in the table on the right.}
\end{figure}

 Despite their different dimensions, each MWPC in MD has the same internal structure, as shown in Fig.\ref{fig:3}. Anode planes are centered inside a 5 mm gas gaps and are formed by gold-plated tungsten wires 30 $\mu$m diameter, with 2 mm spacing. The cathodes are made of FR4 fiberglass plates with two-sided 35 $\mu$m thick copper coating. In regions R1--R3 the cathodes have an additional gold coating of about 100 nm. Adjacent gaps are separated by panels made of honeycomb or rigid polyurethane foam, which provide precise gap alignment over the whole chamber area.

\par The most irradiated MWPCs, operated in region R2 of now removed station M1, integrated already $\sim$0.6 C/cm of charge per unit length of wire over the past nine years, from 2010 until 2018. During this period, the MUON MWPCs did not show any gain reduction or other apparent symptoms of reduced performance in terms of efficiency or time resolution \cite{5, 7}. A very limited fraction of gaps, affected with high currents due to Malter-like effect \cite{8} are well under control, with a developed and successfully applied method of treatment in situ and did not show yet any visible impact on efficiency of the MD \cite{9}. 

\par The absence of aging phenomena in the MD is a key argument for a possible massive reuse of the present MWPCs as a scenario at U2. The expected deposited charge after 50 fb$^{-1}$ of integrated luminosity, which LHCb plans to collect during RUN4 and RUN5 \cite{10} in the regions proposed for instrumentation with MWPCs for U2, is much less than the maximum charge in M1R2 chambers ($\sim$0.6 C/cm of wire), already collected during RUN1 and RUN2, see Tab.\ref{tab:1}. But obviously, it is mandatory to perform additional studies on aging of MWPCs for U2 (500 fb$^{-1}$) conditions, which should include a direct inspection of chambers removed from the most irradiated regions of the MD, such as M1R2, making use of modern tools for microscopic and chemical analysis. In addition, an irradiation campaign aiming at reproducing visible ageing effects in controlled conditions should start in the near future, in such a way to better understand the operational limits of the LHCb chambers.

\begin{table}[htbp]
\centering
\caption{\label{tab:1} Expected average deposited charge (C/cm of wire) after 50 fb$^{-1}$ at U1 in the most irradiated chamber of each station and region of the Muon System. Regions in bold blue are proposed for instrumentation with MWPCs at U2.}
\smallskip
\begin{tabular}{|c|*{4}{|c}|}
\hline
{\bf Station} & {\bf R1} & {\bf R2} & {\bf R3} & {\bf R4} \\
\hline
M2 & 0.67 & 0.42 & \textcolor{blue}{\bf 0.10} & \textcolor{blue}{\bf 0.02} \\
M3 & 0.17 & 0.08 & \textcolor{blue}{\bf 0.02} & \textcolor{blue}{\bf 0.01} \\
M4 & 0.22 & \textcolor{blue}{\bf 0.06} & \textcolor{blue}{\bf 0.01} & \textcolor{blue}{\bf 0.004} \\
M5 & 0.15 & \textcolor{blue}{\bf 0.03} & \textcolor{blue}{\bf 0.01} & \textcolor{blue}{\bf 0.003} \\
\hline
\end{tabular}
\end{table}

\section{Inefficiency and background}

\par Inefficiency in the MD is induced mostly by the dead-time of FEE and by the rate of ghost hits from accidental crossings, in the regions where the reconstructed hits are obtained by crossing large area X- and Y-strips. Especially important is the impact on the M2 station, where the rate can reach 0.6 MHz/cm$^2$ in the most illuminated chamber of region R1, with an efficiency drop as high as 25$\%$ very close to the beam pipe already at U1 conditions. In the dead-time estimate, a 30$\%$ mitigation effect has been included, which accounts for the installation during LS2 of a new shielding in front of the inner region of M2, composed of a new hadron calorimeter (HCAL) beam plug, with a reduced clearance with respect to the beam pipe, and a tungsten shielding installed in place of the innermost cells of the HCAL \cite{11}.

\par Inefficiency from dead-time is proportional to the input rate on FEE channels and could be reduced only by increasing the granularity of the readout pads which is limited by the cluster size of collecting charge.

\par
\begin{figure}[htbp]
\centering 
\includegraphics[width=0.9\textwidth]{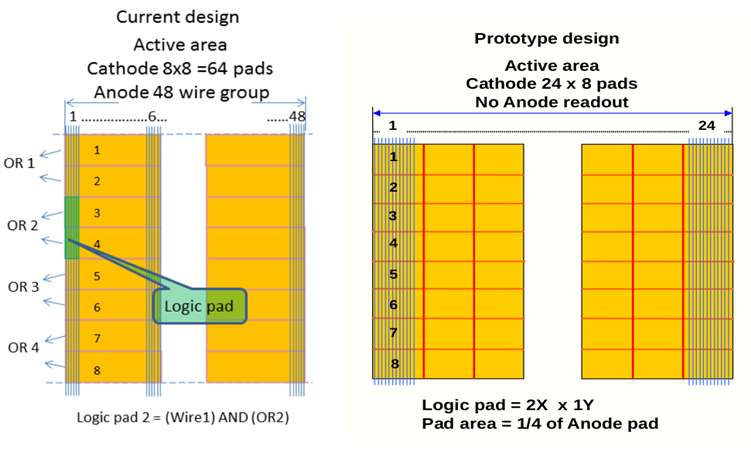}
\caption{\label{fig:4} Present design of R2 chamber in M2 and M3 stations with mixed readout (on the left). The crossing of the physical cathode pads (pad matrix of 8$\times$8) and the 48 physical anode strips forms the logical pads. Simplified design of new high granularity pad chamber with only cathode readout pads (pad matrix of 24$\times$8) is show on the right.}
\end{figure}

 The probability of ghost hits is determined by the rate of output signals from FEE channels where signals are OR-ed in the bi-gaps and also in the intermediate boards (IB) of the off-detector electronics (ODE), forming logical X- and Y-strips. Now the IBs are applied in most of the regions of the Muon System. Moreover, very inner and most occupied regions R1 and R2 of stations M2 and M3, form logical strips on the level of chambers by OR-ing signals from thin in X-plane but long in Y-plane anode pads and wide in X-plane cathode pads, both having quite large readout surface, see Fig.\ref{fig:4} on the left. This construction has been chosen for the initial design of MD on purpose to minimize the amount of FEE channels and accordingly reduce the cost of FEE. Given what's written above, a mitigation of inefficiency from ghosts, keeping the detector unchanged in general, could be reached only by removal IBs to increase granularity of logical strips and by replacing the double readout MWPC with higher granularity pad detectors, see an example for M2R2 chambers on Fig.\ref{fig:4} on the right, which are under production now \cite{12}.

\par An increasing of the readout granularity of X and Y strips is already being realized now at LS2 with the removal of IBs in regions R2, R3 and R4 of station M2 and in region R4 of station M5 \cite{13} (where hit rates are dominated by interactions with LHC materials placed behind of the LHCb detector). The installation of MWPC with full pad readout could already take place during RUN3, depending on the availability of the chambers. The above tricks will reduce loss of the di-muon events, such as ${B}{}^0_s$ $\rightarrow$ $\mu$$^+$$\mu$$^-$, down from $\sim$10$\%$ to $\sim$4.5$\%$ at U1 conditions.

\par
\begin{figure}[htbp]
\centering 
\includegraphics[width=0.8\textwidth]{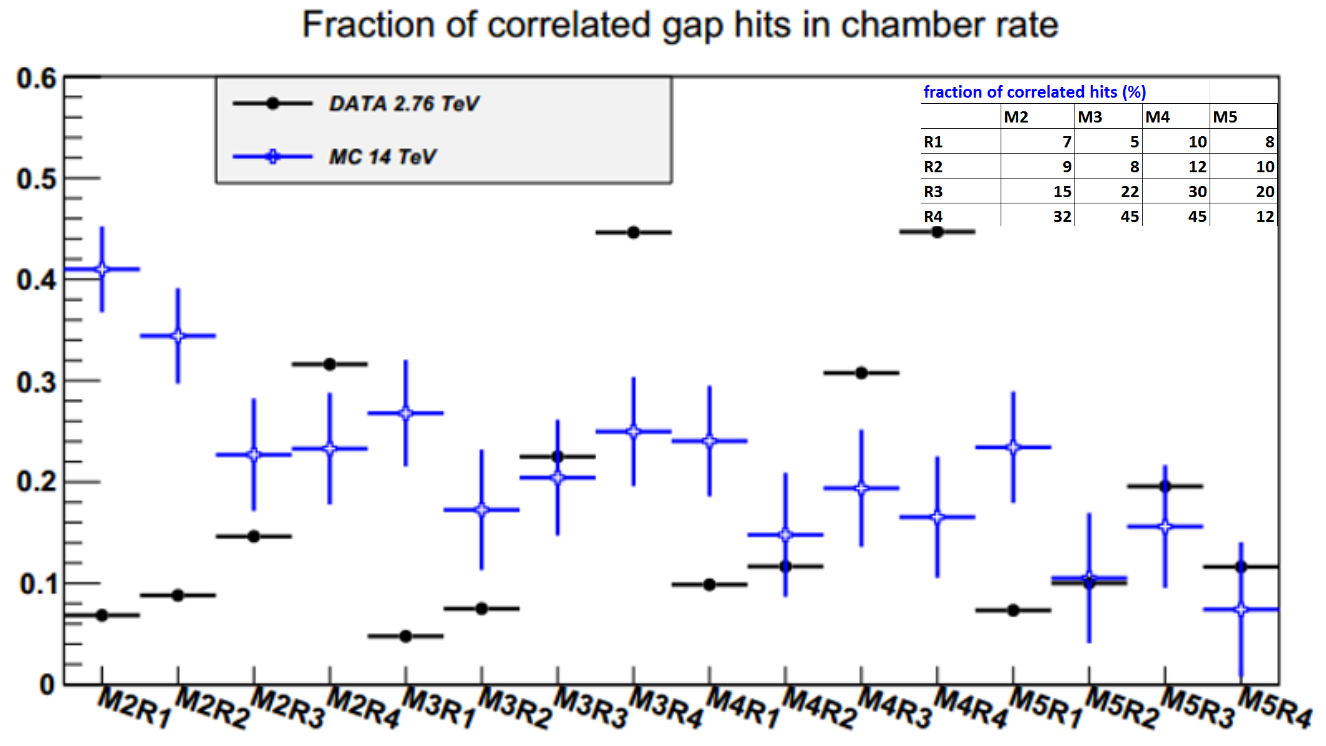}
\caption{\label{fig:5} Average fraction of correlated hits in each of the 16 regions of M2--M5 stations. A summary table is shown in right corner of the plot.}
\end{figure}

 Looking forward on background conditions at U2, where estimations were based mostly on direct measurement of the rates taken in 2012 (see Fig.\ref{fig:5}), clearly show that hits are dominated by low energy background (LEB) particles, hitting only a single gap out of a bi-gap, the so-called uncorrelated hits. According to MC simulation, the most significant fraction of LEB is associated with thermal neutrons produced in hadronic cascades in the calorimeters, in the muon filters or in accelerator parts. They produce low-energy electrons by (n,$\gamma$) reaction and subsequent Compton-scattering or by the photoelectric effect in the detector materials, which can generate signals in only one of the detector gaps.

\par Taking into account that now, during the LS2, LHCb becomes essentially a new detector with new material budget, everything should be rechecked when LHC will restart.

\par Looking on the fraction of uncorrelated hits (which exceeds 90$\%$ in the very inner regions of most of the stations), one can clearly see an obvious features which could be implemented in the new generation of FEE for an upgraded Muon System.

\par First of all, the readout pads in bi-gaps which are now OR-ed at input of the FEE should be separated. It will double the number of FEE channels but mitigate the inefficiency due to dead-time by 30--40$\%$, from LEB events.

\begin{figure}[htbp]
\centering 
\includegraphics[width=0.6\textwidth]{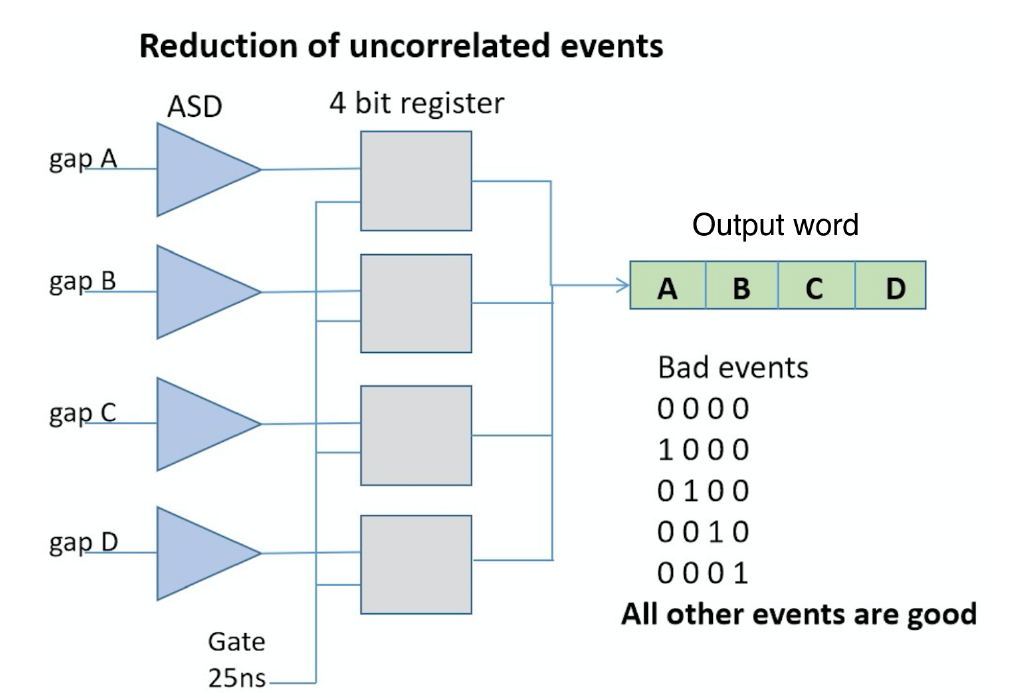}
\caption{\label{fig:6} Scheme for an improved MWPC front-end, with single gap readout and a common gate to reduce the contribution from LEB.}
\end{figure}

\par Second step is the implementation on the level of FEE of a specific logic, see Fig.\ref{fig:6}, which requires at least two hits in four pads in the bunch crossing (BX) window. It will reject most of the events from LEB and will reduce the output rate of FEE by a factor 3 or more.

\section{Proposal with MWPCs for U2}

\par The key parameters to design the new detector are the rate capability, the total instrumented area and the granularity. Understanding these parameters and optimizing the detector design requires of course detailed studies and a proper simulation. While waiting for this, the maximum rate per unit area at the input of each detector gap, see Tab.\ref{tab:2}, has been obtained by scaling to 2$\times$10$^{34}$ cm$^{-2}$s$^{-1}$ the extrapolations prepared for RUN3 \cite{10}, and taking into account fractions of penetrating particles and LEB. The shielding reduction factors on the inner regions of M2 station were included based on estimated from simulation. As a result, regions R1 and R2, $\approx$ 25 m$^2$ of total instrumented surface, will have maximum input rates in the range of several tens of kHz/cm$^2$ to one MHz/cm$^2$, while regions R3 and R4, $\approx$ 365 m$^2$, will have rates up to $\approx$ 10 kHz/cm$^2$.

\begin{table}[htbp]
\centering
\caption{\label{tab:2} Maximum input rate expected at 2$\times$10$^{34}$ cm$^{-2}$s$^{-1}$, total instrumented surface and present logical pad dimension for each station/region of the MD.}
\smallskip
\begin{tabular}{|c|*{3}{|c}|}
\hline
{\bf Region} & {\bf Max rate [kHz/cm$^2$]} & {\bf Total area [m$^2$]} & {\bf Logical pad [cm$^2$]} \\
\hline
M2R1 & 998 & 0.9 & 0.63 $\times$ 3.1  \\
M2R2 & 98 & 3.6 & 1.25 $\times$ 6.3  \\
M2R3 & 13 & 14.4 & 2.5 $\times$ 12.5  \\
M2R4 & 10 & 57.6 & 5 $\times$ 25  \\
\hline
M3R1 & 575 & 1.0 & 0.67 $\times$ 3.4  \\
M3R2 & 72 & 4.2 & 1.35 $\times$ 6.8  \\
M3R3 & 8 & 16.8 & 2.7 $\times$ 13.5  \\
M3R4 & 3 & 67.4 & 5.4 $\times$ 27  \\
\hline
M4R1 & 211 & 1.2 & 2.9 $\times$ 3.6  \\
M4R2 & 30 & 4.9 & 5.8 $\times$ 7.3  \\
M4R3 & 5 & 19.3 & 11.6 $\times$ 14.5  \\
M4R4 & 2 & 77.4 & 23.1 $\times$ 29  \\
\hline
M5R1 & 179 & 1.4 & 3.1 $\times$ 3.9  \\
M5R2 & 20 & 5.5 & 6.2 $\times$ 7.7  \\
M5R3 & 4 & 22.2 & 12.4 $\times$ 15.5  \\
M5R4 & 2 & 88.7 & 24.8 $\times$ 30.9  \\
\hline
\end{tabular}
\end{table}

\par A similar exercise has been performed later on, with rate on FEE channels from data set taken at the end of RUN2 in 2018. This was a special measurement dedicated on the investigation of performance of the LHCb detectors to run at U1 conditions at five times higher instantaneous luminosity. Data were normalized to the maximum filling of the LHC ring with protons bunches (2808 bunches) at nominal luminosity for RUN2 and then extrapolated to the 50 times higher one expected at U2 conditions.

\par To complete the exercise, set of factors reducing the rates on readout pads were included:

\par 1. Effect of new beam plugs for M2 and HCAL and replacement of few inner cells in HCAL by tungsten slabs \cite{11}. The reduction of rates estimated from simulation, is: factor 0.5 in M2R1, 0.75 in M2R2, 0.75 in M3R1 and 0.9 in M3R2. Both pieces to be installed during LS2 and a real effect of this shielding will be checked with data already during RUN3.

\par 2. Replacement HCAL by the Iron wall \cite{14}, which may happen already in 2025-26 during LS3. From simulations, it should reduce the rates in M2 by factor 0.58 in R1, by 0.31 in R2 and by factor 0.36 in R3 and R4 but in D-column of R4 region, due to geometrical size of the wall, which is not enough to shield some outer peripheral areas of the station.

\par 3. Replacement of M2R1, M2R2 and M3R1 MWPCs by pad chambers with higher (by factor 4) granularity, during RUN3 \cite{12}. Reduction factors are 0.5 in M2R1 and 0.375 in M2R2 and in M3R1 are expected. These include effect of cluster size measured in a prototype of new M2R2 MWPC and extrapolated to dimensions of pads in new M2R1 and M3R1 chambers. Measurements on a prototype of M2R1 MWPC have been delayed due to COVID-19 quarantine.

\par 4. Following the idea of new FEE, with separation of the readout pads from bi-gaps to single gaps \cite{3}, factors of input rate reduction were applied, according the measured fraction of penetrated particles (see Fig.\ref{fig:5}) with a safety factor of 2. This was done for all regions in MD but R4 regions of M2--M4 stations, where pads in bi-gaps are galvanically OR-ed inside chambers and there is no technical way to separate them. 

\par Final results are well consistent with previous extrapolation to U2 conditions, see Tab.\ref{tab:2}, which was based on old measurement of input rates performed in 2012 and was already used in the TDR for extrapolation of the rates at U1 \cite{10}.

\par
\begin{figure}[htbp]
\centering 
\includegraphics[width=0.95\textwidth]{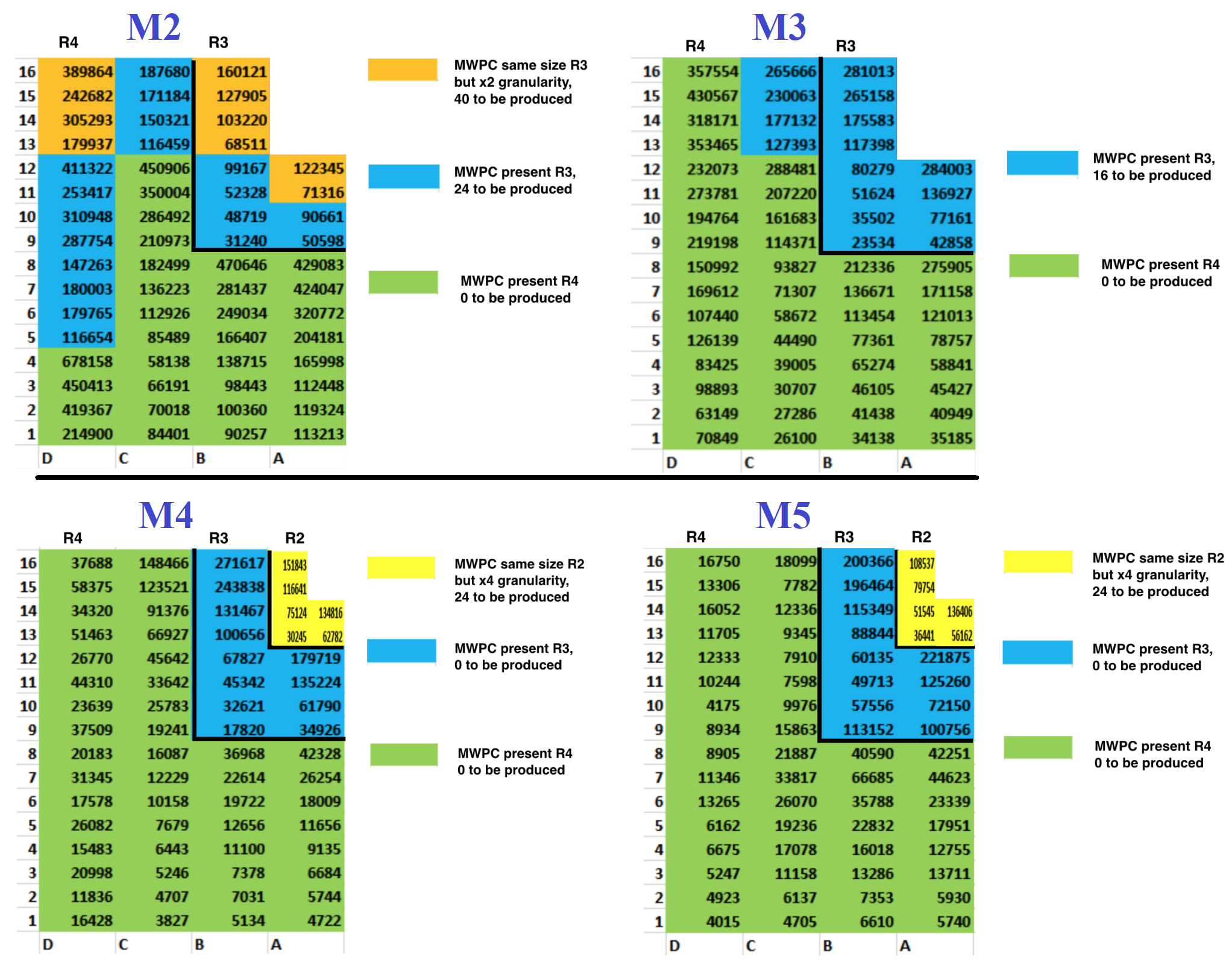}
\caption{\label{fig:7} Regions R3/R4 of M2-M3 and R2/R3/R4 of M4-M5 stations bottom quadrant with the expected maximum rate (Hz) at the input of the FEE channels of each chamber. A colour code indicates the different chamber types. The total number of chambers to be produced is also indicated in the legend.}
\end{figure}

 At the end of this exercise, a safety limit of $\sim$500 kHz as a maximum allowed input rate on FEE channel in the Muon System has been applied to obtain the results. It guarantees that MD can successfully operate with negligible impact in detection inefficiency with a 100 ns dead-time of FEE. Resulting, 10 out of 16 regions of MD were accepted for possible instrumentation either with present MWPCs or with new ones with higher readout pads granularity \cite{3}. In Fig.\ref{fig:7} is reported the maximum rate, at U2 conditions, expected in each chamber of a quadrant and for each Muon station. 

\par A summary of the number of chambers to be reused and produced is given in Tab.\ref{tab:3}, accompanied by the related number of FEE channels. In total, 128 chambers have to be produced, 13$\%$ of the total, out of which 40 have the same design as present R3 chambers, and 88 have higher granularity.

\begin{table}[htbp]
\centering
\caption{\label{tab:3} Number of chambers (CMB) to be reused/produced (w/o spares) for the regions where MWPCs are proposed, and corresponding number of FEE channels.}
\smallskip
\begin{tabular}{|c|*{3}{|c}|}
\hline
{\bf Station/Region} & {\bf CMB Reuse} & {\bf CMB New} & {\bf FEE channels} \\
\hline
M2R3 & 24 & 24 & 58368 \\
M2R4 & 152 & 40 & 6144 \\
\hline
M3R3 & 48 & 0 & 24576 \\
M3R4 & 176 & 16 & 8448 \\
\hline
M4R2 & 0 & 24 & 18432 \\
M4R3 & 48 & 0 & 9216  \\
M4R4 & 192 & 0 & 9216 \\
\hline
M5R2 & 0 & 24 & 18432 \\
M5R3 & 48 & 0 & 9216 \\
M5R4 & 192 & 0 & 18432 \\
\hline
Total & 880 & 128 & 180480 \\
\hline
\end{tabular}
\end{table}

\section{Test setups}

\par
\begin{figure}[htbp]
\centering 
\includegraphics[width=0.8\textwidth]{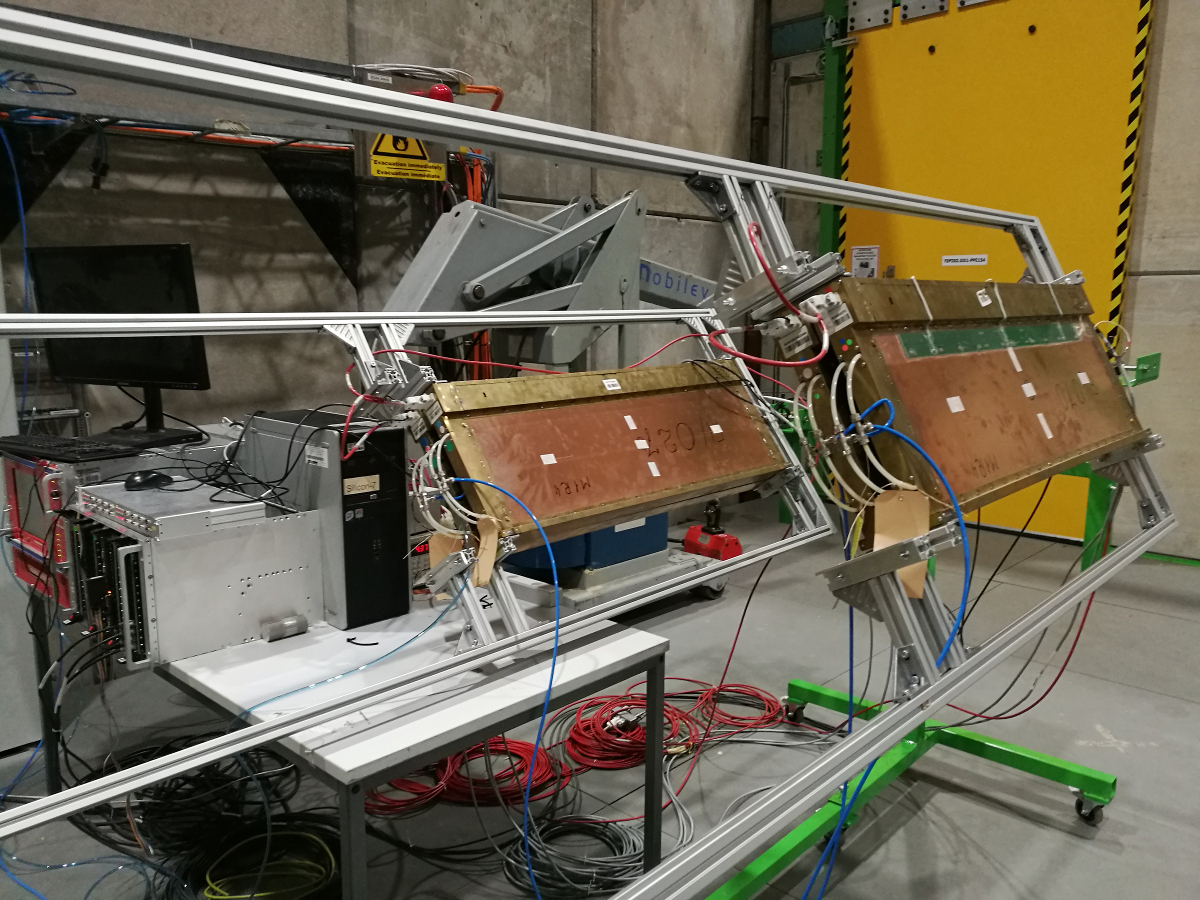}
\caption{\label{fig:8} The setup dedicated for testing of the detectors and electronics at high rate conditions on GIF++. It includes a muon trigger based on MWPCs removed from M1R4. For the muon tracking, are used two sets of two M1R4 chambers each, which are visible on the photo.}
\end{figure}

 One of the significant advantages of this proposal is the availability of test setups for design of both: new high granularity pad MWPCs and a new generation of FEE. One setup has been constructed at Petersburg Nuclear Physics Institute (PNPI) especially for testing with proton beam most of the important parameters (cluster size, time resolution, etc.) of the new pad chambers for inner regions of M2 and M3 stations. The second one was recently built for Gamma Irradiation Facility (GIF++) at CERN \cite{15} and it is dedicated to study detectors of any types and also FEE at high rate conditions, equipped with a very high intensity (16.65 TBq) of the Cs-137 radioactive source. The setup includes a muon trigger, see Fig.\ref{fig:8}, based on a set of MWPCs removed from M1 station and used for efficient track reconstruction of either: incident cosmic muons or provided by a 100 GeV muon beam, which is also available on GIF++. Good tracking of muons at very high irradiation levels from gamma source is possible due to the efficient rejection of not penetrating particles, like conversion electrons (very similar to LEB conditions in the Muon System) when MWPCs are operated in coincidence. In addition, GIF++ is also available for long term aging study of MWPC which is mandatory for U2, and where MUON group already has a fully equipped infrastructure (electronics, gas system, HV, etc.), which is necessary for such investigation (see Fig.\ref{fig:8}).

\section{Conclusion}

\par The LHCb MUON Detector was successfully operated during 10 years of RUN1 and RUN2 data taking with a detection efficiency close to 100$\%$. During this period there was no evidence of any ageing effect in MWPCs, even in the most irradiated area of the Muon System. This very good performance allows to consider an option based on a massive ($\sim$96$\%$ of the area) instrumentation of new MD either with present MWPCs or with higher granularity pad MWPCs ($\sim$15$\%$ from total), provided that the input rate will be compatible with expected abilities of new generation of FEE. The rest of Muon System could be instrumented with new type of detectors, in particular $\mu$-RWELLs.

\par This option has obvious advantages like an acceptable cost in the frame of the expected budget for U2, keeping most of the existing infrastructure unchanged. Other advantages are the experience of the MD group with proportional chambers and the presence of production sites available within the MUON Collaboration, equipped with performant factories for mass-production of MWPCs and also setups for the design and test of both, new pad MWPCs and new FEE. Of course, extensive additional aging studies of MWPCs are mandatory, as well as  dedicated simulations and early measurements of the rates at U1 conditions.

\par Last but not least, all of what reported above clearly indicates that the era of MWPCs as very robust and cheap detectors, which were successfully used in dozens of famous experiments, is not at the end, and proportional chambers can still be efficiently used in many new big and ambitious future experiments, committed to operate in very aggressive conditions in terms of irradiation and occupancy.


\end{document}